\documentstyle[12pt]{article}
    \date{}
\begin{document}
    
    \title{Binding of Quarks and the $\pi N$ $\sigma$-Term.}
    
    \author{L. Ya. Glozman}
    \maketitle
    
    \centerline{\it  Institute for Theoretical Physics, 
    University of Graz,}
    \centerline{\it Universit\"atsplatz 5, A-8010 Graz, Austria}
    
    \setcounter{page} {0}
    \vspace{1cm}
    
    \centerline{\bf Abstract}
    \vspace{0.5cm}
    
    It is shown that the binding effect that is associated with the short      
    range part of the Goldstone boson
    exchange interaction between constituent quarks provides a good
    description of the $\pi N$ $\sigma$-term.\\

    PACS numbers 12.39.Fe, 12.39.-x, 12.38.Lg, 14.20.Dh
    
    \bigskip
    \bigskip

    \bigskip
    \bigskip
    \noindent
    \small
    
    Preprint UNIGRAZ-UTP 09-08-96\\
    
     hep-ph/ 9608283 \\
    
    \newpage
    {\baselineskip=0.7cm
    
    The pion-nucleon $\sigma$-term \cite{CHENG1}
    
    $$\sigma_{\pi N} = \frac {1}{2} (m_u^0 + m_d^0) 
    < N | \bar{u}u +  \bar{d}d | N >, \eqno (1)$$
    
    \noindent
    is a measure of the explicit chiral symmetry breaking effects
    in the nucleon. Here $m_u^0$ and $m_d^0$ stand for the current
    quark masses. Its experimental value may be extracted from 
    pion-nucleon scattering data, the most recent result being \cite{SA}
    
    $$\sigma_{\pi N} = 45 \pm 10 ~ MeV. \eqno (2) $$
    
    Clearly that any successful model of the nucleon should be
    able to explain this empirical value.
    The additive quark ansatz within naive constituent quark model 
    as well as in the extended Nambu-Jona-Lasinio
    model  \cite{BER, VOGL}
    leads to a much smaller value for $\sigma_{\pi N}$. 
    This indicates that some essential piece of physics
    is absent within the additive quark ansatz.
    The aim of
    this letter is to show that
    the effects responsible
    for the binding of the quarks in the nucleon 
    are of crucial importance for the explanation
    of the emperical value of $\sigma_{\pi N}$.\\
    
    The view that in the low-energy regime, i.e. beyond the spontaneous
    breaking of chiral symmetry, light and strange baryons can be viewed
    as  systems of three constituent quarks which interact by the
    exchange of Goldstone bosons (pseudoscalar mesons) and which are subject
    to confinement \cite {GLO2} is becoming rather compelling \cite{GPP}.
    Such an interaction between the light quarks in heavy baryons
    containing one heavy quark is important for understanding the
    spectra of the heavy flavor hyperons as well \cite{GR}.
    The creation of the quark-antiquark sea in the nonperturbative regime
    via the coupling of the Goldstone bosons to the valence quarks
    also resolves some well-known problems related to the spin and
    flavor content of the nucleon that appear in naive constituent quark and
    parton models \cite {QUIGG, CHENG2, ANSELM}. Below we show that the
    contribution to $\sigma_{\pi N}$ that arises from the short-range
    part of Goldstone
    boson exchange (GBE) between the constituent quarks is  crucial
    for the explanation of its empirical value.\\
    
    The pion-nucleon $\sigma$-term can be evaluated via the Feynman -
    Hellmann theorem \cite{HELL,FEY} as:

    $$\sigma_{\pi N} = {\hat{m}}^0 ( \frac {\partial M_N}{\partial m_u^0}
    + \frac {\partial M_N}{\partial m_d^0}), \eqno (3)$$
    
    \noindent
    where ${\hat{m}}^0$ stands for the average 
    value of the current $u$ and $d$
    quarks, $ {\hat{m}}^0 = \frac {1}{2} (m_u^0 + m_d^0)$. 
    Within the constituent
    quark model with chiral dynamics \cite{GLO2,GPP}, the nucleon
    mass consists of four terms:
    
    $$ M_N = \sum_{k=1}^3 m_k + < N | H_{kin} | N > + < N | H_{conf} | N >
    + < N | H_\chi | N >, \eqno (4) $$
    
    \noindent
    where the second, third and fourth terms are contributions
    from the kinetic energy of the constituent quarks, confining interaction,
    and the GBE interaction between constituent quarks,
    respectively. Thus in order to evaluate (3) one needs an explicit
    dependence of each term in (4) on the current quark masses.\\
    
    The constituent mass $m_i$ includes  the current quark mass
    value $m_i^0$ as well as a dynamical part $m_i^D$:
    
    $$ m_i = m_i^0 + m_i^D. \eqno (5)$$
    
    \noindent
    The latter  
    appears from the spontaneous chiral symmetry breaking. 
    In the chiral limit, $m_i^0=0$, 
    the constituent quark mass is determined by  the quark 
    condensates $< \bar{q}q >$, 
    which, in turn,
    are defined as the closed quark loops. Thus the dynamical part in (5)
    is in principle dependent on the full mass $m_i$ and equation (5)
    becomes a gap (Schwinger-Dyson) equation. Obviously, no solution of this
    equation that takes into account full gluodynamics is presently
    available.
    Nevertheless, near the chiral limit, $m_u^0 = m_d^0 =0$, the dynamical 
    part
    in (5) is  weakly dependent on the current quark masses  $m_u^0$
    and  $m_d^0$. This feature is well seen  from the solution of
    the gap equation in the Nambu and Jona-Lasinio model well beyond
    the critical value of the coupling constant \cite{VOGL}. Thus, for
    a rough estimate one can use (near the chiral limit)
    
    $$\frac {\partial m_i}{\partial m_j^0} \simeq \delta_{ij}, \eqno(6)$$
    
    \noindent
    where  $i,j = u$ or $d$.\\
    
    The kinetic term in (4) exibits $ m^{-1}$ dependence on the constituent
    quark mass, and thus
    
    $$ < N | \frac {\partial H_{kin}}{\partial m_u^0} + 
    \frac {\partial H_{kin}}{\partial m_d^0} | N > =
    - \frac {1}{m} < N | H_{kin} | N >. \eqno(7)$$
    
    \noindent
    Here and in what follows we assume for simplicity
    equal masses for the constituent $u$ and $d$ quarks, $ m_u = m_d = m$.\\
    
    Assuming that the confining interaction is determined by the gluodynamics,
    one concludes that the confining term in (4) does not contribute
    to $\sigma_{\pi N}$. This also follows from the fact that the
    effective confining interaction between the constituent quarks 
    does not depend on their masses.\\
    
    The GBE interaction between the constituent quarks is proportional
    to $m^{-2}$ \cite {GLO2,GPP}, and thus
    
    $$ < N | \frac {\partial H_\chi}{\partial m_u^0} + 
    \frac {\partial H_\chi}{\partial m_d^0} | N > =
    - \frac {2}{m} < N | H_\chi | N >. \eqno(8)$$
    
    \noindent
    The repulsive contribution to the nucleon mass 
    of the Yukawa tail of the quark-quark interaction,
    $\sim  \mu^2\exp{(-\mu r)}/r$, where $\mu$ is the meson mass, is very
    small \cite{GPP}, and hence the dependence
    of $H_\chi$ on $m_u^0$ and $m_d^0$ via meson mass $\mu$ in the Yukawa         
    tail is not important and is neglected in (8). The crucial attractive
    contribution to the nucleon mass from the GBE comes from its
    short-range part which is $\mu$-independent and 
    has opposite sign relative to the Yukawa tail. It is this opposite sign
    which is the key to the explanation of the baryon spectrum \cite{GLO2}.
    In the chiral limit the long-range Yukawa tail vanishes, while
    the short-range part of GBE remains intact.\\

    Using in what follows the average value for the light quark masses
    ${\hat {m}}^0 = 7$ MeV \cite{LEUT} and a standard value 
    $m= 340$ MeV for the 
    constituent quark mass (which is suggested by the nucleon magnetic
    moments and which is used in the parametrization of the qq potential
    in \cite {GPP}), the pion-nucleon
    sigma-term can now be estimated (3). For that we shall use 
    the numerical values
    of  $< N | H_{kin} | N >$ and  $< N | H_\chi | N >$ developed
    in three-body Faddeev calculations in ref. \cite {GPP}. With
    a parametrization of the GBE given therein, one has:

    $$ < N | H_{kin} | N > = 844 ~ MeV,$$

    $$ < N | H_\chi | N > = -1130 ~ MeV,$$

    $$ < N | H_{conf} | N > = 204 ~ MeV. \eqno(9) $$

    \noindent
    The sum of all these terms plus $3m = 1020$ MeV gives just the
    nucleon mass.\\

    One then obtains:

    $$\sigma_{\pi N} \simeq 3 \times 7 - \frac {844}{340} \times 7
    + 2\frac { 1130}{340} \times 7 = 50.1 ~ MeV \eqno(10)$$

    \noindent
    This result is in good agreement with the empirical value (2)
     as well as with the recent lattice QCD calculation \cite{LIU},
    where   $\sigma_{\pi N} = 47 - 53$ MeV. \\

    One cannot insist, however, that it is the value ${\hat {m}}^0 = 7$ MeV
    which is responsible for the result (10). It would be so if the assumption
    (6) were exact. In fact the numerical value of the  $\sigma_{\pi N}$
    is determined by the products 
    ${\hat {m}}^0\frac {\partial m_i}{\partial m_j^0}$,
    thus it is better to say that the result (10) is achieved with
    ${\hat {m}}^0( \frac {\partial m_i}{\partial m_i^0}+ 
    \frac {\partial m_i}{\partial m_j^0} )= 7$ MeV. Hence, 
    the smaller values for the current quark mass could  also be
    compatible with the empirical value of $\sigma_{\pi N}$ provided that
    the dependence of $m^D_i$ on $m^0_j$ in Eq. (5) is essential.
    This question could be answered only when we have achieved a better
    understanding of a microscopical nature of the constituent quark.
    This uncertainty does not affect however the main conclusion that
    is discussed below.\\

    Usually the nucleon is considered as a system of three weakly
    interacting constituent quarks as $M_N \simeq 3m$. This is
    not an adequate view. It can be seen from the $\Delta - N$
    mass splitting that a difference of the expectation values
    of the spin-spin forces between the quarks for $N$ and $\Delta$
    should be of order 300 MeV. It is clear that the contribution
    of the spin-spin interaction to the nucleon has to be much
    bigger than the difference above. The big binding effect from the
    GBE is compensated mostly by the large kinetic energy as well as by the
    confining interaction. Such a compensation is well seen in (9).
    However, there is no such a compensation between 
    $ < N | H_\chi | N >$ and $ < N | H_{kin} | N > $ contributions
    in  $\sigma_{\pi N}$ in Eq. (10) as the weight factor 
    of $ < N | H_\chi | N >$
     is twice bigger than the corresponding weight factor 
    of the kinetic energy contribution 
    as it is seen in Eqs. (7) and (8).
    Thus it is the big
    absolute value of $ < N | H_\chi | N >$ which is crucial for the
    explanation of the  $\sigma_{\pi N}$ within the constituent quark
    model.\\

    \
    
    \end{document}